**Measurements of minimum entropy at K-shell photoelectric effect provides the radius and shape of the proton.**

Jiménez Edward[1], Recalde Nicolas[2], Jimenez Esteban[3]
[1] Faculty of Chemical Engineering, Central University of Ecuador, Quito, Ecuador.
[2] University of South Carolina, USA.  [3] Université Paul Sabatier, FRANCE.
Correspondance, E-mail: ehjimenez@uce.edu.ec

**Abstract**

Constructive resonances with minimum entropy are derived from measurements of photoelectric effect cross sections at atomic K-shell thresholds. A resonance region with optimal constructive interference is given by a principal wave length λ of the order of Bohr atom radius. Our study shows that the proton shape is not a sphere but it has an elliptical volumetric shape with two equal axes and the other different.

Resonance waves allowed us the possibility to measure both proton radius and shape through an interference term. This last, was a necessary condition in order to have an effective cross section maximum at threshold. The minimum entropy means minimum shape deformation and it was found to be (0.830 ± 0.015) fm and the average proton radius was found to be (0.851 ± 0.201) fm.

# Introduction

The so called proton radius puzzle was born when electron scattering experiments or atomic spectroscopy gave different value for the radius of the proton when compared to a muonic experiment. In the last experiment, the lamb shift was measured when the muon orbits a proton allowing a proton radius measurement[1]. Other experiments included a Deuterium spectroscopy and muonic Helium. It was found in the former that the deuteron radius is smaller when measured in muonic deuterium compared to the average value using electronic deuterium. There are also plans to simultaneously measure the scattering of electrons and muons at Muon Proton Scattering Experiment (MUSE)[1,2].

The electrons in the atomic K-shell orbit much closer to the nucleus than other shells. At low energy (up to 0.116 MeV) is well known that photoelectric effect is more important than other effects for most of elements of periodic table[3]. We can see different peaks of photoelectric absorption being the strongest ones due to electrons located at K-shell. Somehow the system formed by electron, photon and nucleus interacts strongly at K-shell when compared to other shells. The NIST data base at National Institute of Standards and Technology[4] has tabulated the photoelectric peak cross sections for every element of periodic table. Resonance is a very general phenomenon in nature but many times we ignore its existence. It can be discovered when the energy in a system changes at particular rate[5]. Using NIST information and entropy principles we develop a model that call for atomic resonances for the electron, proton and photon system[6]. Atomic resonances where completely identified through an interference term that became part of the model and allowed us to measured proton radius. Our model predicts that the proton is not an sphere but an ellipse and it is in agreement with other experiments done by other laboratories[1,2].

## Theory

Resonance phenomena can be seen at low energy when x-rays interact with periodic table atoms. This is true when we analyze attenuation peaks produced from low energy radiation interaction with matter[5,7,8,9].

If we consider that resonance is a threshold phenomenon we can find a specific wave length $\lambda$ where maximum energy is transferred from one photon to other electron or proton[10,11]. In our analysis, low energy X-rays interacts with atomic structure as a system and produce resonances. We have seen minimum increases of energy that produce more than 250% of the photoelectric effect cross section ($\sigma_m \gg \sigma_{m-1}$) for a time t=m, and for each atom of the periodic table. When we consider resonance threshold frequencies, we found that the maximum variation of the cross section is located where the energy variation is minimum $Min(E_m-E_{m-1})$ or equal to zero during the same time interval, as follows:

$$Min(E_m - E_{m-1}) \Leftrightarrow Max(\sigma_m - \sigma_{m-1}) \qquad (1)$$

The resonance phenomena is an optimal process, in the sense that entropy is minimized. During resonance system entropy should be minimum and can be measured when huge increases of atomic cross sections are detected. This last is the result of interference phenomena between the atomic nucleus, X-rays and atomic electrons[12].

In order to have a clear understanding how these process occurs, we are going to explain each one of the events. Those events are ordered and optimized naturally in such a way that a maximum value of the cross section is obtained for every atom between: $11 \leq Z \leq 92$.

E1 **Interaction of matter with radiation**: The X-rays having a wave length $\lambda$, interacts constructively with a given material and as a result of this a resonance is produced[12,13].

E2 **Constructive interference**: There are constructive interferences among the X-rays and later on with the atomic nucleus or electron. This occur only if $r_a \approx r_n + \lambda$ where $r_a$ = atomic radius and $r_n$ = nucleus radius.

E3 **The resonance maximize total atomic cross section of the system**. This resonance comes mainly from the nucleus since it has the largest interaction probability when compared with the electron: $p_n > p_e$, where $p_n$ and $p_e$ are the nucleon and electron probability respectively[14,15].

E4 **The optimal cross section of the atomic nucleus depends on the proton radius $r_p$**. The proton radius will be measured indirectly through the atomic total cross section which is a result of the interference between the nucleus (but always with a specific nucleon), X-rays and electron[16].

$$\frac{p_n+p_e}{p_n}(\sigma_2 - \sigma_1) = Max(\sigma_m - \sigma_{m-1}) = 8\pi r_p \lambda \qquad (2)$$

The process is described through the events E1, E2, E3 and E4 and is only possible if and only if exists both a simultaneous constructive interference and atomic total cross section resonance[10-16].

This quantum phenomena implies an optimal and ordered development of each of the physical events, namely E1, E2, E3 and E4. Therefore, the principle that governs the whole process is called minimum entropy. The proton radius will be then theoretically calculated by minimizing the entropy[17,18].

## Definitions.

The atom will be characterized by the atomic radius $r_a$, number of protons Z = A-N, number of electrons Z and number of neutrons N.

**Cross Section.**

According to NIST, current tabulations of μ/ρ rely heavily on theoretical values for the total cross section per atom, $\sigma_{tot}$, which is related to μ/ρ by the following equation:

$$\frac{\mu}{\rho} = \sigma_{tot}/uA \qquad (3)$$

In (eq. 3), u (= 1.660 540 2 × 10$^{-24}$) is the atomic mass unit (1/12 of the mass of an atom of the nuclide 12C)[4].

The attenuation coefficient, photon interaction cross sections and related quantities are functions of the photon energy. The total cross section can be written as the sum over contributions from the principal photon interactions,

$$\sigma_{tot} = \sigma_{pe} + \sigma_{coh} + \sigma_{incoh} + \sigma_{pair} + \sigma_{trip} + \sigma_{ph.n} \qquad (4)$$

Where $\sigma_{pe}$ is the atomic photo effect cross section, $\sigma_{coh}$ and $\sigma_{incoh}$ are the coherent (Rayleigh) and the incoherent (Compton) scattering cross sections, respectively, $\sigma_{pair}$ and $\sigma_{trip}$ are the cross sections for electron-positron production in the fields of the nucleus and of the atomic electrons, respectively, and $\sigma_{ph.n}$ is the photonuclear cross section[3,4].

We use data of NIST for elements Z=11 to Z=92 and photon energies 1.0721E-03 MeV to 1.16E-01 MeV, and have been calculated according to:

(5)

$$\frac{\mu}{\rho} = (\sigma_{pe} + \sigma_{coh} + \sigma_{incoh} + \sigma_{pair} + \sigma_{trip} + \sigma_{ph.n})/uA$$

The probability to find a proton inside the atom and more specifically inside the nucleus is given by $P_n$:

$$P_n = \frac{\sigma_n}{\sigma_a} = \frac{1.2^2 A^{2/3}}{r_a^2}. \tag{6}$$

The probability to find at least one electron inside the atom is represented by Bohr's radius and is defined by:

$$P_e = \frac{\sigma_e}{\sigma_a - \sigma_n} = \frac{r_e^2}{r_a^2 - 1.2^2 A^{2/3}}. \tag{7}$$

Where:

$r_a$ is the atomic radius,

$r_e$ is the electron radius,

$r_n$ is the nucleus radius,

$r_p$ is the proton radius.

## Resonance region.

A resonance region is created in a natural way at the K-shell between the nucleus and the electrons at S-level. The condition for the photons to enter in the resonance region is given by $r_a \geq r_n + \lambda$. This resonance region give us a new way to understand the photoelectric effect. There is experimental evidence of the existence of resonance at K-level due to photoelectric effect, represented by the resonance cross section provided by NIST for each atom. In the present work we focus on the resonance effects but not on the origin of resonance region.

**Shannon Entropy.**

Shannon's entropy S measures the information of a system and is given by:

$$S = \frac{\sigma_n}{\sigma_a - \sigma_n} \ln\left(\frac{\sigma_a - \sigma_n}{\sigma_n}\right) + \frac{\sigma_n}{\sigma_a} \ln\left(\frac{\sigma_a}{\sigma_n}\right). \tag{8}$$

In order to have minimum entropy[19,20,21] every part of the atom should have a shape and structure optimal in a way that the whole system works as one. The equation (8) only depends on the nucleus radius. The minimum nucleus radius will be used $r_n^* = r_p A^{1/3}$ instead of traditional nucleus radius given by $r_n^* = 1.2\ A^{1/3}$. This new equation can be obtained from the following volumetric relation:

$$\frac{4}{3}\pi r_n^3 = A \frac{4}{3}\pi r_p^{*3} \tag{8A}$$

The equation (8A) will be used in order to minimize the entropy.

**Theorem 1. The atom is a minimal entropy system.**

$$\min\left\{S = \frac{A^{2/3}r_p^2}{r_a^2}\ln\left(\frac{r_a^2}{A^{2/3}r_p^2}\right) + \frac{Zr_e^2}{r_a^2 - A^{2/3}r_p^2}\ln\left(\frac{r_a^2 - A^{2/3}r_p^2}{Zr_e^2}\right)\right\}$$

$$\frac{A^{2/3}r_p^2}{r_a^2} + \frac{Zr_e^2}{r_a^2 - A^{2/3}r_p^2} < 1 \qquad (9)$$

The problem is solved using the KKT (Karush-Kund-Tucker) method[15,19]. The Lagrangian is built in a traditional way L = S + μ (p$_n$+p$_e$-1) with the constraint μ(p$_n$+p$_e$-1)= 0.

After solving the Lagrangian, dL/dr$_p$ = 0 we obtain μ > 0 as follows:

$$\mu =$$

$$\frac{-r_a^2 2r_p^2 r_e^2 A^{2/3}\ln\left(\frac{-A^{2/3}r_p^2 + r_a^2}{Z}\right) - 2r_a^2 r_e^2 A^{2/3}\ln\left(\frac{1}{r_e^2}\right) + 2r_p r_e^2 A^{2/3}Zr_a^2}{2rr_e^2 A^{2/3}Z + 2rA^{2/3}\left(A^{2/3}r_p^2 - r_a^2\right)^2} +$$

$$\frac{-2r_p A^{2/3}\left(A^{2/3}r_p^2 - r_a^2\right)^2 \ln\left(\frac{r_a^2}{rA^{2/3}}\right) + 2r_p A^{2/3}\left(A^{2/3}r_p^2 - r_a^2\right)^2}{2r_p r_e^2 A^{2/3}Z + 2r_p A^{2/3}\left(A^{2/3}r_p^2 - r_a^2\right)^2}. \qquad (10)$$

Since μ > 0, we should have

$$\frac{A^{2/3}r_p^2}{r_a^2} + \frac{Zr_e^2}{r_a^2 - A^{2/3}r_p^2} = 1. \qquad (11)$$

This implies that the minimum of the entropy has solution since μ is larger than zero and equation (9) has solution. In other words, r$_p$ exists in the interval from 0.83 fm to 0.88 fm for every one of the atoms. In summary the entropy is minimum for the minimum radius.

❶

Equation (11) depends on the radius of electron and nucleon. From this equation we can solve for the proton radius for each element of the periodic table. We then obtain the theoretical limit of the proton radius according to Figure 3.

Once the limits of proton radius have been established we found that they are in agreement to experimental values. Therefore, we can find the proton radius value of minimum entropy by using the following equation.

$$\min\left\{S = \frac{A^{2/3}r_p^2}{r_a^2}\ln\left(\frac{r_a^2}{A^{2/3}r_p^2}\right) + \frac{Zr_e^2}{r_a^2 - A^{2/3}r_p^2}\ln\left(\frac{r_a^2 - A^{2/3}r_p^2}{Zr_e^2}\right)\right\} \qquad (12)$$
$$0{,}83 fm < r_p < 0{,}88\ fm$$

**Theorem 2. Resonance region. The resonance cross section is produced by interference between the atomic nucleus and the incoming X-rays inside the resonance region, where the boundaries are the surface of the atomic nucleus and K-shell.**

The cross section of the atomic nucleus is given by:

$$\sigma_{r_n} = 4\pi r_n^2 = 4\pi A^{2/3} r_n^2 \qquad (13)$$

The photon cross section at K-shell depends on the wave length and the shape of the atomic nucleus:

$$\sigma_{r_n + \lambda} = 4\pi(r_n + \lambda)^2 \qquad (14)$$

Subtracting the cross sections (13) and (14) we have:

$$\sigma_\lambda = \sigma_{r_n+\lambda} - \sigma_{r_n} = 4\pi(2r_n\lambda + \lambda^2) = 4\pi\left(2r_p\lambda + 2(r_n - r_p)\lambda + \lambda^2\right) \qquad (15)$$

The resonance is produced by interactions between the X-rays, the K-shell electrons and the atomic nucleus. The cross sections corresponding to the nucleus is weighted by probability $p_n$ and should have a simple dependence of an interference term. This last depends on the proton radius $r_p$ or the difference between the nucleus and proton radius ($r_n$ - $r_p$) according to the following relation:

$$\frac{p_n + p_e}{p_n}(\sigma_2 - \sigma_1) = Max(\sigma_m - \sigma_{m-1}) = 4\pi(2r_p\lambda) \qquad (16)$$

$$\frac{p_n + p_e}{p_n}(\sigma_2 - \sigma_1) = Max(\sigma_m - \sigma_{m-1}) = 4\pi(2(r_n - r_p)\lambda) \qquad (17)$$

We note that left hand side of equations (16) and (17) should have a factor larger than one due to resonance. The unique factor that holds this requirement is ($p_n$ + $p_e$ )/ $p_n$.

Due to physical reasons the term that should be taken is $4\pi(2r_p\lambda)$ because this is only that is in boundary of the resonance region.

The reason that proton shape should be elliptical is because we can only measure one axis of the ellipse and current experimental values are in the range of 0.83 fm to 0.88 fm.

It should be noted that the X-rays interaction is with a given proton located at the nucleus boundary. Therefore, the semi-empirical version of the equation (16), which has been validated using experimental data from NIST and theoretical foundations, is then given by:

$$\frac{\frac{r_e^2}{r_a^2-1.2^2A^{2/3}}+\frac{1.2^2A^{2/3}}{r_a^2}}{\frac{1.2^2A^{2/3}}{r_a^2}}(\sigma_2-\sigma_1)=8\pi b\lambda \tag{18}$$

From equation (18) we obtain that proton dimensions are given by the two principal axis of an ellipse where the proton radius is defined by the relation $r_p^2$ = ab.

❷

We have then shown that we can verify equation (18) using NIST data for each of the elements of the periodic table. The results for ellipse axe values and proton radius can be seen in Table 1.

## Discussion of Results.

1.- The shape and dimensions of the proton are dynamic and correspond to an ellipse. The proton is deformed and increases with the atomic weight A, due to nuclear force. Using the minimum entropy theory we can calculate the optimal dimension of the proton radius and the conditions for the photon to be trapped in the resonance region corresponding to K-shell. Once in this region the photon interacts with the electron or with one of the nucleons. In this paper we give a calculation method for the proton radius as a function of the resonance cross section. However, in a similar way we can obtain the electron radius since the low energy X-rays are confined between to boundaries corresponding to electrons at K-shell and protons in the nucleus.

2.- The proton shape corresponds to an ellipse with two equal length axis and the other different. From the experimental cross section we can obtain two axis, the larger and the smaller. From the experiment we only obtain one axis which is in the range of 1.3 fm ≤ b ≤ 0.59 fm, where 1.3 fm corresponds to sodium Z=11 and 0.59 corresponds to Uranium Z=92. We can calculate the second axis by using the formula $r_p^2$ = ab.

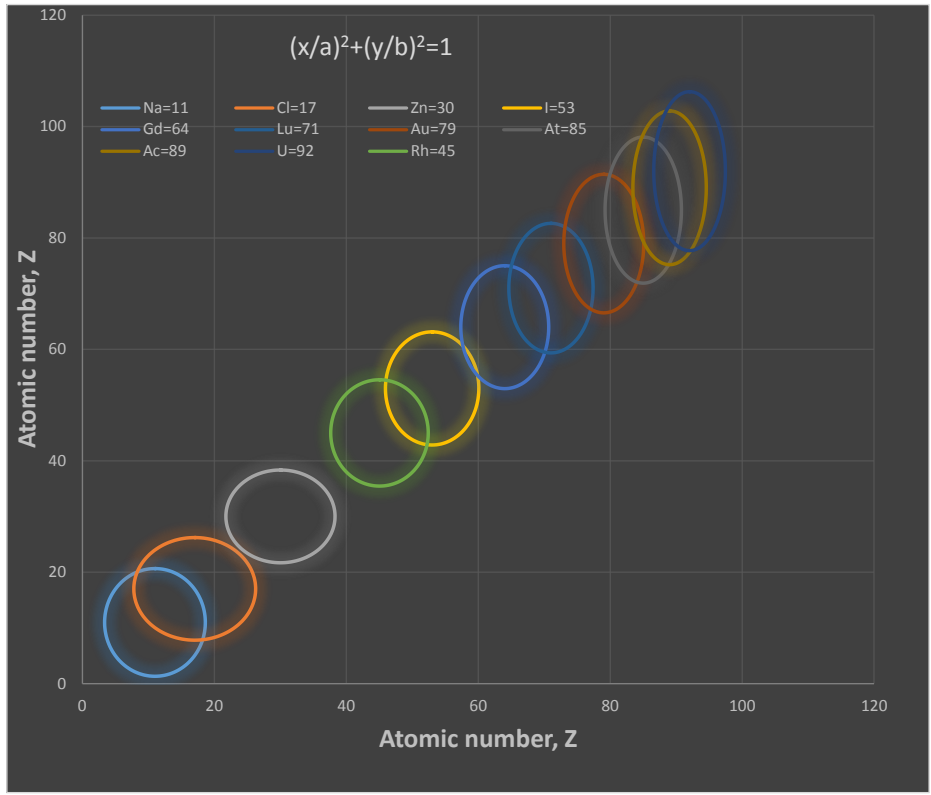

Figure 1. Evolution of size and shape for proton inside the atom nucleus. We can see different revolution ellipses for every atom for Z=11 to Z=92.

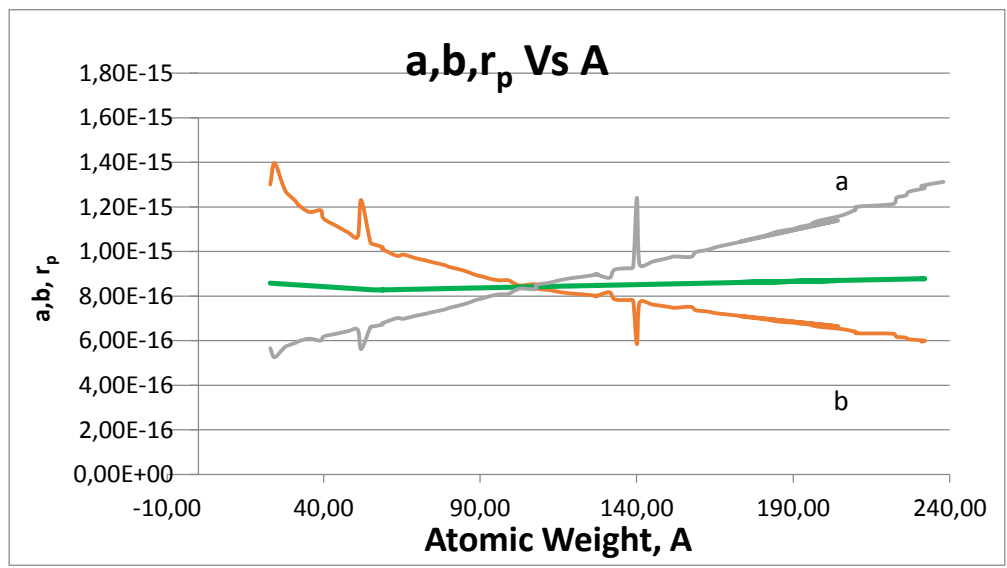

Figure 2. Variation of nucleus radius (green), principal and secondary ellipse axes (orange, silver).

3.- Considering minimal deformation, the most probable value for the radius of the proton is obtained by averaging the measured value for one of axis of the ellipse. The value obtained was surprisingly 0.851 fm.

4.- This model allow us to explain the mass absorption curves for low energy X-rays and help us interpret the photoelectric effect in terms of interference wave properties. The question about why the K-shell has more probability than the more external ones can be understood in terms of the effects of the nucleus in the resonance and the same for the electron. Moreover, the photons that come into the resonance region have a small probability to produce a primary photoelectric effect and for this reason the secondary effect is predominant.

The solid rigid shape of the nucleus is completely known in terms of its radius with a precision given by:

$$r_n = (1.25 \pm 0.051) A^{1/3} \tag{19}$$

The equation (19) gives the functional dependence the nucleus radius $r_n$ in terms of atomic number A. If we consider that the nucleons have similar relationship we then have:

$$r_p = (0{,}855 \pm 0{,}0348) \tag{20}$$

However, from available experimental data we found that the range of variation of proton radius is well below to the corresponding nuclear radius.

$$r_p = (0{,}855 \pm 0{,}025) \tag{21}$$

According to our experimental values from Table 1, the value of the ellipse major axis, with one standard deviation, is given by:

$$b = (0{,}851 \pm 0.199) \tag{22}$$

The scale of variation of our data is given in fm, which is within range of other laboratories.

**5.-** The entropy is minimum for the minimum proton radius for different atoms of periodic table. See Figure 2 and Table 2.

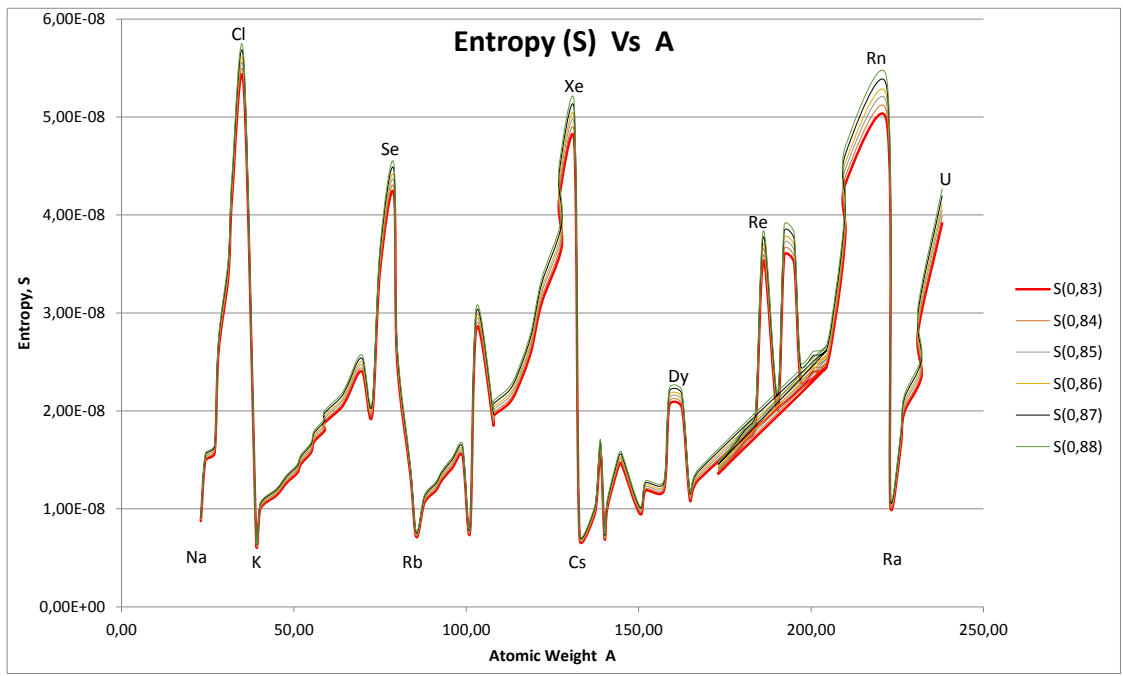

Figure 3. The figure shows minimum entropy values for different elements of the periodic table where 11 ≤ Z ≤ 92. We can see that minimum entropy corresponds to a value of the proton radius equal to 0.83 fm.

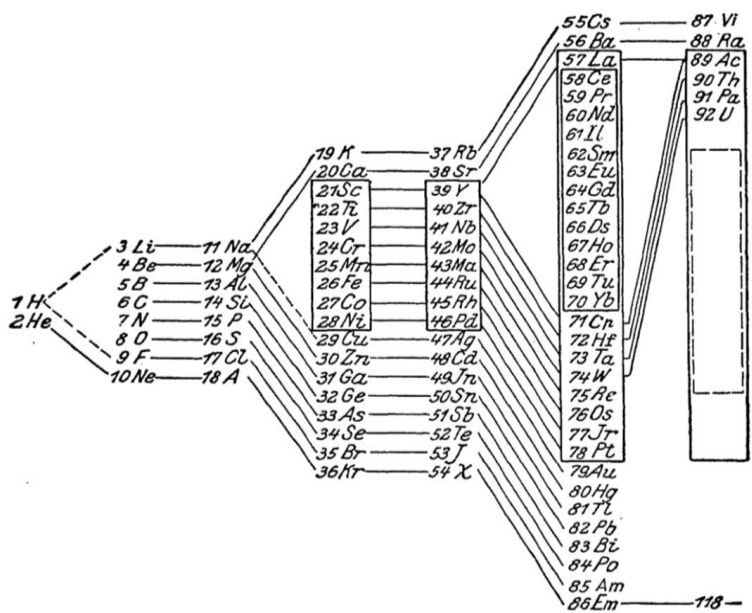

Figure 4. This figure shows a periodic system in relation to closed shells according to Max Born[7], Note the close relationship with figure 3, which was obtained from entropy principles only. This strongly suggest the idea of closed resonance regions inside the atoms.

5.- The overlapping of the cross section for the nucleons is minimum as we can see from the evolution of the probability interaction against the nucleus. This is a function of the atomic weight, Figure 5. This is true when we consider the empty space in the atomic nucleus. This last can be seen from the difference between the minimal nucleon radius and the known proton radius, given by equation (23).

$$\frac{r_p - r_p^*}{r_p} = 0{,}445785 \tag{23}$$

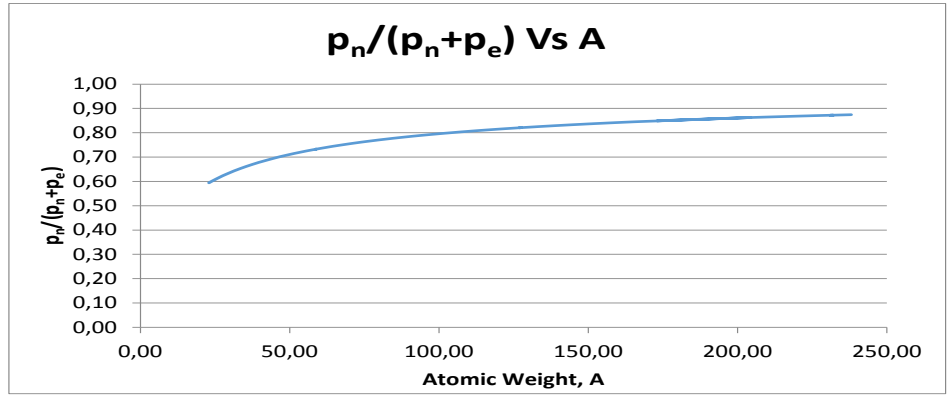

Figure 5. Nucleus probability as a function of Atomic weight A.

**Author contributions**

E. Jimenez and N. Recalde conceived the theoretical model. Esteban Jimenez processed information. E. Jimenez and N. Recalde wrote the manuscript.

**Methods**

We compile all official information from PML NIST division for elements from Z=1 to Z=92 for energies ranged (10E-3 MeV to 1.16E-1 MeV) corresponding to K-shell. Using the equations 8, 9, 10, 16 we obtained the final results reported in Table 1 and 2. Note that photoelectric effect by resonance only start to appear from Z=11 to Z=92.

## Bibliography.

1. Strauch Steffen, Fifth Workshop of the APS Topical Group on Hadronic Physics, April 2013, Denver, CO, USA.

2. Strauch Steffen, PANIC14, Hamburg, Germany, August 24 - 29, 2014 abstract.

3. Wong Samuel, Introductory Nuclear Physics, Second Edition, Wiley-VCH Verlag GmbH & C0. KGaA, Germany, 2004.

4	J.H Hubbell, S.M. Seltzer, X-Ray Mass Attenuation Coefficients, Radiation Division, https://www,nist.gov/pml/x-ray-mass-attenuation-coefficients, PML, NIST, 2017.

5	A. C. G. Mitchell, M.W. Zemansky, Resonance radiation and exited atoms, Cambridge, 2009.

6.	Gordon Christopher, Quantum Information Theory and the foundation of Quantum Mechanics, Oxford University, Thesis submitted for the Degree of Doctor of Philosophy, London, England 2004.

7.	Born Max, Atomic Physics, Second Edition, Blackie & Son Limited, England 1937.

8.	Jackson John David, Classic Electrodynamics, John Wiley & Sons, 2004.

9	L.D. Landau and E.M. Lifschitz, Mechanics, Nauka Publishers, Moscow, 1988.

10.	Cohen Tannudji Claude, Quantum Mechanics, Wiley-VCH, 1992.

11	R. Pohl et al., Nature 466, 213 (2010).

12	C. Cohen, D. Guery, Advances in Atomic Physics, WSPC, 2011.

13	J.S. Briggs 1, A.M. Lane, The effect of atomic binding on a nuclear resonance, Physics Letters B, Vol 106, Issue 6, Pages 436-438, 1981

14	D. Griffits, Introduction to elementary particles, Second edition, Wiley-VCH, 2008

15.	W. Karush, Isoperimetric Problems and Index Theorems in the Calculus of Variations, Doctoral Dissertation, Department of Mathematics, University of Chicago, 1942.

16	G. A. Bliss, The problem of Lagrange, American Journal of Mathematics 52 (1930), 673–744.

17	G. A. Bliss, Lectures on the Calculus of Variations, University of Chicago Press, Chicago, 1946.

18	M. R. Hestenes, Calculus of Variations and Optimal Control Theory, John Wiley & Sons, New York, 1966.

19	W. Karush, Mathematical Programming, Man-Computer Search and System Control. Technical Report SP-828, System Development Corporation, Santa Monica,D Calif., 1962.

20	H. W. Kuhn and A. W. Tucker, Nonlinear programming, in (J. Neyman, ed.) Proceedings of the Second Berkeley Symposium on Mathematical Statistics and Probability, University of California Press, Berkeley, 1951, pp. 481–492.

21	L. A. Pars, An Introduction to the Calculus of Variations, John Wiley & Sons, New York, 1962.

| ELEMENT | Z | N | A | $E_2=E_1=E$ | $r_e$ | $r_a$ | $\sigma_2$ | $\sigma_1$ | $\lambda$ | $(\sigma_2-\sigma_1)/\sigma_1$ | b | a | $r_p$ | $r_n^*$ | $r_n$ | S | $p_n$ | $p_e$ | $1-p_n-p_e$ | $p_n/(p_n+p_e)$ |
|---|---|---|---|---|---|---|---|---|---|---|---|---|---|---|---|---|---|---|---|---|
| | | | A | MeV | m | m | cm² | cm² | m | % | m | m | m | m | m | | | | | |
| Na | 11 | 12 | 22,99 | 1,072E-03 | 2,82E-15 | 1,900E-10 | 2,46E-19 | 2,07E-20 | 1,1580E-09 | 1085% | 1,30E-15 | 5,66E-16 | 8,58E-16 | 8,97E-15 | 9,20E-15 | 1,25E-08 | 3,23E-10 | 2,2029E-10 | 1,0000E+00 | 0,59 |
| Mg | 12 | 12 | 24,31 | 1,305E-03 | 2,82E-15 | 1,450E-10 | 2,20E-19 | 1,83E-20 | 9,5135E-10 | 1102% | 1,40E-15 | 5,26E-16 | 8,57E-16 | 9,47E-15 | 9,72E-15 | 2,14E-08 | 5,75E-10 | 3,7824E-10 | 1,0000E+00 | 0,60 |
| Al | 13 | 14 | 26,98 | 1,560E-03 | 2,82E-15 | 1,431E-10 | 1,77E-19 | 1,62E-20 | 7,9605E-10 | 993% | 1,30E-15 | 5,62E-16 | 8,54E-16 | 1,05E-14 | 1,08E-14 | 2,28E-08 | 6,33E-10 | 3,8835E-10 | 1,0000E+00 | 0,62 |
| Si | 14 | 14 | 28,09 | 1,840E-03 | 2,82E-15 | 1,110E-10 | 1,49E-19 | 1,44E-20 | 6,7474E-10 | 932% | 1,27E-15 | 5,75E-16 | 8,53E-16 | 1,09E-14 | 1,12E-14 | 3,77E-08 | 1,08E-09 | 6,4543E-10 | 1,0000E+00 | 0,63 |
| P | 15 | 16 | 30,97 | 2,150E-03 | 2,82E-15 | 9,800E-11 | 1,27E-19 | 1,28E-20 | 5,7745E-10 | 892% | 1,23E-15 | 5,90E-16 | 8,51E-16 | 1,20E-14 | 1,24E-14 | 4,97E-08 | 1,48E-09 | 8,2803E-10 | 1,0000E+00 | 0,64 |
| S | 16 | 16 | 32,06 | 2,470E-03 | 2,82E-15 | 8,800E-11 | 1,10E-19 | 1,16E-20 | 5,0264E-10 | 854% | 1,21E-15 | 5,98E-16 | 8,50E-16 | 1,24E-14 | 1,28E-14 | 6,19E-08 | 1,88E-09 | 1,0269E-09 | 1,0000E+00 | 0,65 |
| Cl | 17 | 18 | 35,45 | 2,822E-03 | 2,82E-15 | 7,900E-11 | 9,65E-20 | 1,04E-20 | 4,3994E-10 | 827% | 1,18E-15 | 6,09E-16 | 8,47E-16 | 1,36E-14 | 1,42E-14 | 7,92E-08 | 2,49E-09 | 1,2742E-09 | 1,0000E+00 | 0,66 |
| K | 19 | 20 | 39,10 | 3,607E-03 | 2,82E-15 | 2,430E-10 | 7,79E-20 | 8,64E-21 | 3,4416E-10 | 802% | 1,18E-15 | 6,00E-16 | 8,43E-16 | 1,50E-14 | 1,56E-14 | 9,66E-09 | 2,81E-10 | 1,3467E-10 | 1,0000E+00 | 0,68 |
| Ca | 20 | 20 | 40,08 | 4,038E-03 | 2,82E-15 | 1,940E-10 | 6,81E-20 | 7,85E-21 | 3,0745E-10 | 767% | 1,15E-15 | 6,19E-16 | 8,43E-16 | 1,53E-14 | 1,60E-14 | 1,50E-08 | 4,48E-10 | 2,1130E-10 | 1,0000E+00 | 0,68 |
| Sc | 21 | 24 | 44,96 | 4,490E-03 | 2,82E-15 | 1,840E-10 | 6,08E-20 | 7,23E-21 | 2,7651E-10 | 741% | 1,11E-15 | 6,34E-16 | 8,38E-16 | 1,71E-14 | 1,80E-14 | 1,75E-08 | 5,38E-10 | 2,3489E-10 | 1,0000E+00 | 0,70 |
| Ti | 22 | 26 | 47,88 | 4,970E-03 | 2,82E-15 | 1,760E-10 | 5,47E-20 | 6,66E-21 | 2,4980E-10 | 721% | 1,09E-15 | 6,43E-16 | 8,35E-16 | 1,82E-14 | 1,92E-14 | 1,95E-08 | 6,13E-10 | 2,5673E-10 | 1,0000E+00 | 0,70 |
| V | 23 | 28 | 50,94 | 5,465E-03 | 2,82E-15 | 1,710E-10 | 4,97E-20 | 6,16E-21 | 2,2717E-10 | 707% | 1,07E-15 | 6,49E-16 | 8,33E-16 | 1,93E-14 | 2,04E-14 | 2,12E-08 | 6,77E-10 | 2,7196E-10 | 1,0000E+00 | 0,71 |
| Cr | 24 | 28 | 51,99 | 5,989E-03 | 2,82E-15 | 1,660E-10 | 5,16E-20 | 5,68E-21 | 2,0729E-10 | 809% | 1,23E-15 | 5,62E-16 | 8,32E-16 | 1,97E-14 | 2,08E-14 | 2,27E-08 | 7,28E-10 | 2,8859E-10 | 1,0000E+00 | 0,72 |
| Mn | 25 | 30 | 54,94 | 6,540E-03 | 2,82E-15 | 1,610E-10 | 4,12E-20 | 5,29E-21 | 1,8983E-10 | 679% | 1,04E-15 | 6,60E-16 | 8,29E-16 | 2,07E-14 | 2,20E-14 | 2,46E-08 | 8,03E-10 | 3,0679E-10 | 1,0000E+00 | 0,72 |
| Fe | 26 | 30 | 55,85 | 7,110E-03 | 2,82E-15 | 1,558E-10 | 3,78E-20 | 4,93E-21 | 1,7462E-10 | 667% | 1,03E-15 | 6,64E-16 | 8,28E-16 | 2,10E-14 | 2,23E-14 | 2,64E-08 | 8,67E-10 | 3,2761E-10 | 1,0000E+00 | 0,73 |
| Co | 27 | 32 | 58,93 | 7,710E-03 | 2,82E-15 | 1,520E-10 | 3,48E-20 | 4,61E-21 | 1,6103E-10 | 656% | 1,02E-15 | 6,73E-16 | 8,28E-16 | 2,22E-14 | 2,36E-14 | 2,84E-08 | 9,44E-10 | 3,4420E-10 | 1,0000E+00 | 0,73 |
| Ni | 28 | 30 | 58,69 | 8,330E-03 | 2,82E-15 | 1,490E-10 | 3,21E-20 | 4,32E-21 | 1,4904E-10 | 643% | 1,01E-15 | 6,78E-16 | 8,28E-16 | 2,21E-14 | 2,35E-14 | 2,94E-08 | 9,80E-10 | 3,5820E-10 | 1,0000E+00 | 0,73 |
| Cu | 29 | 34 | 63,55 | 8,979E-03 | 2,82E-15 | 1,450E-10 | 2,93E-20 | 4,04E-21 | 1,3827E-10 | 626% | 9,80E-16 | 7,02E-16 | 8,29E-16 | 2,40E-14 | 2,54E-14 | 3,22E-08 | 1,09E-09 | 3,7824E-10 | 1,0000E+00 | 0,74 |
| Zn | 30 | 34 | 65,41 | 9,660E-03 | 2,82E-15 | 1,420E-10 | 2,76E-20 | 3,81E-21 | 1,2852E-10 | 624% | 9,86E-16 | 6,98E-16 | 8,30E-16 | 2,47E-14 | 2,62E-14 | 3,39E-08 | 1,16E-09 | 3,9439E-10 | 1,0000E+00 | 0,75 |
| Ga | 31 | 38 | 69,72 | 1,037E-02 | 2,82E-15 | 1,350E-10 | 2,56E-20 | 3,59E-21 | 1,1976E-10 | 613% | 9,69E-16 | 7,13E-16 | 8,31E-16 | 2,63E-14 | 2,79E-14 | 3,85E-08 | 1,34E-09 | 4,3635E-10 | 1,0000E+00 | 0,75 |
| Ge | 32 | 41 | 72,64 | 1,110E-02 | 2,82E-15 | 1,520E-10 | 2,39E-20 | 3,39E-21 | 1,1185E-10 | 605% | 9,60E-16 | 7,21E-16 | 8,32E-16 | 2,75E-14 | 2,91E-14 | 3,13E-08 | 1,09E-09 | 3,4420E-10 | 1,0000E+00 | 0,76 |
| As | 33 | 42 | 74,92 | 1,187E-02 | 2,82E-15 | 1,140E-10 | 2,23E-20 | 3,21E-21 | 1,0462E-10 | 595% | 9,52E-16 | 7,29E-16 | 8,33E-16 | 2,84E-14 | 3,00E-14 | 5,50E-08 | 1,97E-09 | 6,1191E-10 | 1,0000E+00 | 0,76 |
| Se | 34 | 44 | 78,96 | 1,266E-02 | 2,82E-15 | 1,030E-10 | 2,08E-20 | 3,04E-21 | 9,8083E-11 | 586% | 9,38E-16 | 7,41E-16 | 8,34E-16 | 2,99E-14 | 3,16E-14 | 6,85E-08 | 2,50E-09 | 7,4959E-10 | 1,0000E+00 | 0,77 |
| Br | 35 | 44 | 79,90 | 1,347E-02 | 2,82E-15 | 1,350E-10 | 1,95E-20 | 2,89E-21 | 9,2144E-11 | 576% | 9,32E-16 | 7,47E-16 | 8,34E-16 | 3,03E-14 | 3,20E-14 | 4,11E-08 | 1,47E-09 | 4,3635E-10 | 1,0000E+00 | 0,77 |
| Kr | 36 | 48 | 83,80 | 1,433E-02 | 2,82E-15 | 1,900E-10 | 1,83E-20 | 2,74E-21 | 8,6664E-11 | 566% | 9,19E-16 | 7,59E-16 | 8,35E-16 | 3,18E-14 | 3,35E-14 | 2,19E-08 | 7,64E-10 | 2,2029E-10 | 1,0000E+00 | 0,78 |
| Rb | 37 | 48 | 85,47 | 1,520E-02 | 2,82E-15 | 2,650E-10 | 1,72E-20 | 2,61E-21 | 8,1684E-11 | 558% | 9,11E-16 | 7,66E-16 | 8,36E-16 | 3,25E-14 | 3,42E-14 | 1,17E-08 | 3,98E-10 | 1,1324E-10 | 1,0000E+00 | 0,78 |
| Sr | 38 | 50 | 87,62 | 1,610E-02 | 2,82E-15 | 2,190E-10 | 1,61E-20 | 2,49E-21 | 7,7091E-11 | 546% | 9,00E-16 | 7,77E-16 | 8,36E-16 | 3,33E-14 | 3,50E-14 | 1,71E-08 | 5,92E-10 | 1,6581E-10 | 1,0000E+00 | 0,78 |
| Y | 39 | 50 | 88,91 | 1,704E-02 | 2,82E-15 | 2,120E-10 | 1,52E-20 | 2,38E-21 | 7,2866E-11 | 538% | 8,94E-16 | 7,83E-16 | 8,37E-16 | 3,38E-14 | 3,56E-14 | 1,83E-08 | 6,38E-10 | 1,7694E-10 | 1,0000E+00 | 0,78 |
| Zr | 40 | 51 | 91,22 | 1,800E-02 | 2,82E-15 | 2,060E-10 | 1,43E-20 | 2,27E-21 | 6,8985E-11 | 531% | 8,86E-16 | 7,91E-16 | 8,37E-16 | 3,47E-14 | 3,65E-14 | 1,96E-08 | 6,88E-10 | 1,8740E-10 | 1,0000E+00 | 0,79 |
| Nb | 41 | 52 | 92,91 | 1,899E-02 | 2,82E-15 | 1,980E-10 | 1,36E-20 | 2,17E-21 | 6,5392E-11 | 523% | 8,79E-16 | 7,99E-16 | 8,38E-16 | 3,54E-14 | 3,72E-14 | 2,13E-08 | 7,53E-10 | 2,0285E-10 | 1,0000E+00 | 0,79 |
| Mo | 42 | 54 | 95,94 | 2,000E-02 | 2,82E-15 | 1,900E-10 | 1,28E-20 | 2,09E-21 | 6,2079E-11 | 515% | 8,71E-16 | 8,08E-16 | 8,39E-16 | 3,66E-14 | 3,84E-14 | 2,34E-08 | 8,36E-10 | 2,2029E-10 | 1,0000E+00 | 0,79 |
| Tc | 43 | 55 | 98,91 | 2,104E-02 | 2,82E-15 | 1,830E-10 | 1,23E-20 | 2,02E-21 | 5,8996E-11 | 509% | 8,71E-16 | 8,09E-16 | 8,40E-16 | 3,77E-14 | 3,96E-14 | 2,56E-08 | 9,20E-10 | 2,3746E-10 | 1,0000E+00 | 0,79 |
| Ru | 44 | 57 | 101,07 | 2,210E-02 | 2,82E-15 | 2,650E-10 | 1,15E-20 | 1,91E-21 | 5,6177E-11 | 504% | 8,56E-16 | 8,25E-16 | 8,40E-16 | 3,86E-14 | 4,04E-14 | 1,27E-08 | 4,45E-10 | 1,1324E-10 | 1,0000E+00 | 0,80 |
| Rh | 45 | 58 | 102,91 | 2,320E-02 | 2,82E-15 | 1,340E-10 | 1,10E-20 | 1,85E-21 | 5,3514E-11 | 494% | 8,48E-16 | 8,34E-16 | 8,41E-16 | 3,93E-14 | 4,12E-14 | 4,72E-08 | 1,76E-09 | 4,4288E-10 | 1,0000E+00 | 0,80 |
| Pd | 46 | 60 | 107,87 | 2,440E-02 | 2,82E-15 | 1,690E-10 | 1,06E-20 | 1,79E-21 | 5,0882E-11 | 490% | 8,54E-16 | 8,31E-16 | 8,42E-16 | 4,13E-14 | 4,31E-14 | 3,11E-08 | 1,14E-09 | 2,7844E-10 | 1,0000E+00 | 0,80 |
| Ag | 47 | 60 | 107,90 | 2,55E-02 | 2,82E-15 | 1,650E-10 | 9,92E-21 | 1,72E-21 | 4,8660E-11 | 478% | 8,34E-16 | 8,50E-16 | 8,42E-16 | 4,13E-14 | 4,32E-14 | 3,25E-08 | 1,20E-09 | 2,9210E-10 | 1,0000E+00 | 0,80 |
| Cd | 48 | 64 | 112,41 | 2,670E-02 | 2,82E-15 | 1,610E-10 | 9,45E-21 | 1,64E-21 | 4,6499E-11 | 475% | 8,27E-16 | 8,61E-16 | 8,44E-16 | 4,31E-14 | 4,50E-14 | 3,48E-08 | 1,29E-09 | 3,0679E-10 | 1,0000E+00 | 0,81 |
| In | 49 | 64 | 114,80 | 2,790E-02 | 2,82E-15 | 1,560E-10 | 9,02E-21 | 1,59E-21 | 4,4499E-11 | 469% | 8,20E-16 | 8,69E-16 | 8,44E-16 | 4,41E-14 | 4,58E-14 | 3,74E-08 | 1,40E-09 | 3,2678E-10 | 1,0000E+00 | 0,81 |
| Sn | 50 | 69 | 118,71 | 2,920E-02 | 2,82E-15 | 1,450E-10 | 8,59E-21 | 1,53E-21 | 4,2518E-11 | 462% | 8,12E-16 | 8,80E-16 | 8,45E-16 | 4,56E-14 | 4,75E-14 | 4,37E-08 | 1,65E-09 | 3,7824E-10 | 1,0000E+00 | 0,81 |
| Sb | 51 | 70 | 121,80 | 3,049E-02 | 2,82E-15 | 1,330E-10 | 8,24E-21 | 1,48E-21 | 4,0717E-11 | 457% | 8,09E-16 | 8,85E-16 | 8,46E-16 | 4,69E-14 | 4,87E-14 | 5,22E-08 | 2,00E-09 | 4,4957E-10 | 1,0000E+00 | 0,82 |
| Te | 52 | 76 | 127,60 | 3,180E-02 | 2,82E-15 | 1,230E-10 | 7,88E-21 | 1,43E-21 | 3,9041E-11 | 452% | 8,01E-16 | 8,98E-16 | 8,48E-16 | 4,92E-14 | 5,10E-14 | 6,20E-08 | 2,41E-09 | 5,2564E-10 | 1,0000E+00 | 0,82 |
| I | 53 | 74 | 126,90 | 3,317E-02 | 2,82E-15 | 1,150E-10 | 7,55E-21 | 1,38E-21 | 3,7430E-11 | 447% | 7,99E-16 | 8,99E-16 | 8,48E-16 | 4,89E-14 | 5,08E-14 | 7,03E-08 | 2,75E-09 | 6,0132E-10 | 1,0000E+00 | 0,82 |
| Xe | 54 | 77 | 131,30 | 3,560E-02 | 2,82E-15 | 1,080E-10 | 7,24E-21 | 1,34E-21 | 3,4874E-11 | 442% | 8,17E-16 | 8,82E-16 | 8,49E-16 | 5,07E-14 | 5,25E-14 | 8,07E-08 | 3,19E-09 | 6,8179E-10 | 1,0000E+00 | 0,82 |
| Cs | 55 | 78 | 132,90 | 3,590E-02 | 2,82E-15 | 2,980E-10 | 6,93E-21 | 1,29E-21 | 3,4583E-11 | 436% | 7,86E-16 | 9,18E-16 | 8,49E-16 | 5,13E-14 | 5,32E-14 | 1,17E-08 | 4,22E-10 | 8,9550E-11 | 1,0000E+00 | 0,83 |
| Ba | 56 | 81 | 137,30 | 3,740E-02 | 2,82E-15 | 2,530E-10 | 6,66E-21 | 1,25E-21 | 3,3196E-11 | 431% | 7,82E-16 | 9,25E-16 | 8,51E-16 | 5,31E-14 | 5,49E-14 | 1,63E-08 | 5,99E-10 | 1,2424E-10 | 1,0000E+00 | 0,83 |
| La | 57 | 82 | 138,90 | 3,890E-02 | 2,82E-15 | 1,950E-10 | 6,39E-21 | 1,22E-21 | 3,1916E-11 | 426% | 7,78E-16 | 9,32E-16 | 8,51E-16 | 5,37E-14 | 5,56E-14 | 2,69E-08 | 1,02E-09 | 2,0914E-10 | 1,0000E+00 | 0,83 |
| Ce | 58 | 82 | 140,10 | 3,598E-02 | 2,82E-15 | 2,980E-10 | 6,13E-21 | 1,92E-21 | 3,4506E-11 | 219% | 5,84E-16 | 1,24E-15 | 8,52E-16 | 5,42E-14 | 5,60E-14 | 1,20E-08 | 4,37E-10 | 8,9550E-11 | 1,0000E+00 | 0,83 |
| Pr | 59 | 82 | 140,90 | 4,199E-02 | 2,82E-15 | 2,470E-10 | 5,89E-21 | 1,14E-21 | 2,9567E-11 | 415% | 7,69E-16 | 9,43E-16 | 8,52E-16 | 5,46E-14 | 5,64E-14 | 1,73E-08 | 6,39E-10 | 1,3035E-10 | 1,0000E+00 | 0,83 |
| Nd | 60 | 83 | 144,24 | 4,360E-02 | 2,82E-15 | 2,060E-10 | 5,68E-21 | 1,11E-21 | 2,8475E-11 | 411% | 7,66E-16 | 9,49E-16 | 8,53E-16 | 5,59E-14 | 5,77E-14 | 2,47E-08 | 9,33E-10 | 1,8740E-10 | 1,0000E+00 | 0,83 |
| Pm | 61 | 84 | 145,00 | 4,520E-02 | 2,82E-15 | 2,050E-10 | 5,47E-21 | 1,08E-21 | 2,7467E-11 | 406% | 7,62E-16 | 9,55E-16 | 8,53E-16 | 5,62E-14 | 5,80E-14 | 2,50E-08 | 9,46E-10 | 1,8923E-10 | 1,0000E+00 | 0,83 |
| Sm | 62 | 88 | 150,36 | 4,680E-02 | 2,82E-15 | 2,590E-10 | 5,24E-21 | 1,05E-21 | 2,6528E-11 | 399% | 7,52E-16 | 9,72E-16 | 8,55E-16 | 5,84E-14 | 6,01E-14 | 1,63E-08 | 6,07E-10 | 1,1855E-10 | 1,0000E+00 | 0,84 |
| Eu | 63 | 88 | 152,00 | 4,850E-02 | 2,82E-15 | 2,310E-10 | 5,05E-21 | 1,02E-21 | 2,5598E-11 | 394% | 7,48E-16 | 9,78E-16 | 8,55E-16 | 5,91E-14 | 6,08E-14 | 2,04E-08 | 7,69E-10 | 1,4903E-10 | 1,0000E+00 | 0,84 |
| Gd | 64 | 93 | 157,30 | 5,023E-02 | 2,82E-15 | 2,330E-10 | 4,92E-21 | 9,95E-22 | 2,4717E-11 | 394% | 7,52E-16 | 9,76E-16 | 8,57E-16 | 6,12E-14 | 6,29E-14 | 2,05E-08 | 7,73E-10 | 1,4648E-10 | 1,0000E+00 | 0,84 |
| Tb | 65 | 94 | 158,90 | 5,200E-02 | 2,82E-15 | 1,750E-10 | 4,69E-21 | 9,69E-22 | 2,3877E-11 | 384% | 7,37E-16 | 9,97E-16 | 8,57E-16 | 6,19E-14 | 6,36E-14 | 3,55E-08 | 1,38E-09 | 2,5967E-10 | 1,0000E+00 | 0,84 |
| Dy | 66 | 97 | 162,50 | 5,379E-02 | 2,82E-15 | 1,770E-10 | 4,52E-21 | 9,44E-22 | 2,3081E-11 | 379% | 7,31E-16 | 1,01E-15 | 8,58E-16 | 6,34E-14 | 6,50E-14 | 3,52E-08 | 1,37E-09 | 2,5384E-10 | 1,0000E+00 | 0,84 |
| Ho | 67 | 98 | 164,93 | 5,560E-02 | 2,82E-15 | 2,470E-10 | 4,35E-21 | 9,20E-22 | 2,2329E-11 | 373% | 7,24E-16 | 1,02E-15 | 8,59E-16 | 6,44E-14 | 6,60E-14 | 1,88E-08 | 7,10E-10 | 1,3035E-10 | 1,0000E+00 | 0,84 |
| Er | 68 | 99 | 167,26 | 5,749E-02 | 2,82E-15 | 2,260E-10 | 4,21E-21 | 8,98E-22 | 2,1597E-11 | 368% | 7,20E-16 | 1,03E-15 | 8,59E-16 | 6,53E-14 | 6,69E-14 | 2,24E-08 | 8,56E-10 | 1,5570E-10 | 1,0000E+00 | 0,85 |
| Tm | 69 | 122 | 204,38 | 5,940E-02 | 2,82E-15 | 1,700E-10 | 4,07E-21 | 1,06E-21 | 2,0901E-11 | 285% | 6,65E-16 | 1,14E-15 | 8,70E-16 | 8,08E-14 | 8,18E-14 | 4,49E-08 | 1,73E-09 | 2,7517E-10 | 1,0000E+00 | 0,86 |
| Yb | 70 | 103 | 173,05 | 6,133E-02 | 2,82E-15 | 2,220E-10 | 3,92E-21 | 8,53E-22 | 2,0242E-11 | 360% | 7,11E-16 | 1,04E-15 | 8,61E-16 | 6,77E-14 | 6,92E-14 | 2,36E-08 | 9,07E-10 | 1,6136E-10 | 1,0000E+00 | 0,85 |
| Lu | 71 | 104 | 175,00 | 6,331E-02 | 2,82E-15 | 2,170E-10 | 3,79E-21 | 8,35E-22 | 1,9609E-11 | 354% | 7,06E-16 | 1,05E-15 | 8,62E-16 | 6,85E-14 | 7,00E-14 | 2,48E-08 | 9,57E-10 | 1,6888E-10 | 1,0000E+00 | 0,85 |
| Ta | 73 | 108 | 180,90 | 6,742E-02 | 2,82E-15 | 2,000E-10 | 3,54E-21 | 7,97E-22 | 1,8416E-11 | 345% | 6,96E-16 | 1,07E-15 | 8,63E-16 | 7,10E-14 | 7,24E-14 | 2,95E-08 | 1,15E-09 | 1,9881E-10 | 1,0000E+00 | 0,85 |
| W | 74 | 109 | 183,84 | 6,950E-02 | 2,82E-15 | 1,930E-10 | 3,42E-21 | 7,78E-22 | 1,7864E-11 | 339% | 6,89E-16 | 1,08E-15 | 8,64E-16 | 7,22E-14 | 7,35E-14 | 3,18E-08 | 1,25E-09 | 2,1349E-10 | 1,0000E+00 | 0,85 |
| Re | 75 | 110 | 186,21 | 7,170E-02 | 2,82E-15 | 1,370E-10 | 3,31E-21 | 7,61E-22 | 1,7315E-11 | 335% | 6,85E-16 | 1,09E-15 | 8,65E-16 | 7,32E-14 | 7,45E-14 | 6,16E-08 | 2,50E-09 | 4,2370E-10 | 1,0000E+00 | 0,86 |
| Os | 76 | 114 | 190,23 | 7,387E-02 | 2,82E-15 | 1,850E-10 | 3,21E-21 | 7,45E-22 | 1,6807E-11 | 331% | 6,81E-16 | 1,10E-15 | 8,66E-16 | 7,49E-14 | 7,61E-14 | 3,52E-08 | 1,39E-09 | 2,3236E-10 | 1,0000E+00 | 0,86 |
| Ir | 77 | 114 | 192,22 | 7,610E-02 | 2,82E-15 | 1,370E-10 | 3,11E-21 | 7,28E-22 | 1,6314E-11 | 327% | 6,76E-16 | 1,11E-15 | 8,67E-16 | 7,57E-14 | 7,69E-14 | 6,27E-08 | 2,56E-09 | 4,2370E-10 | 1,0000E+00 | 0,86 |
| Pt | 78 | 117 | 195,08 | 7,839E-02 | 2,82E-15 | 1,390E-10 | 3,01E-21 | 7,13E-22 | 1,5837E-11 | 323% | 6,68E-16 | 1,12E-15 | 8,68E-16 | 7,69E-14 | 7,80E-14 | 6,15E-08 | 2,51E-09 | 4,1159E-10 | 1,0000E+00 | 0,86 |
| Au | 79 | 118 | 197,00 | 8,070E-02 | 2,82E-15 | 1,740E-10 | 2,91E-21 | 7,00E-22 | 1,5384E-11 | 316% | 6,65E-16 | 1,13E-15 | 8,68E-16 | 7,77E-14 | 7,88E-14 | 4,03E-08 | 1,61E-09 | 2,6266E-10 | 1,0000E+00 | 0,86 |
| Hg | 80 | 121 | 200,59 | 8,310E-02 | 2,82E-15 | 1,710E-10 | 2,82E-21 | 6,86E-22 | 1,4940E-11 | 311% | 6,59E-16 | 1,15E-15 | 8,69E-16 | 7,92E-14 | 8,02E-14 | 4,20E-08 | 1,69E-09 | 2,7196E-10 | 1,0000E+00 | 0,86 |
| Tl | 81 | 122 | 204,40 | 8,550E-02 | 2,82E-15 | 1,700E-10 | 2,73E-21 | 6,72E-22 | 1,4521E-11 | 307% | 6,54E-16 | 1,16E-15 | 8,70E-16 | 8,09E-14 | 8,18E-14 | 4,29E-08 | 1,73E-09 | 2,7517E-10 | 1,0000E+00 | 0,86 |
| Pb | 82 | 125 | 207,20 | 8,800E-02 | 2,82E-15 | 1,540E-10 | 2,64E-21 | 6,57E-22 | 1,4107E-11 | 302% | 6,48E-16 | 1,17E-15 | 8,71E-16 | 8,20E-14 | 8,29E-14 | 5,22E-08 | 2,13E-09 | 3,3532E-10 | 1,0000E+00 | 0,86 |
| Bi | 83 | 126 | 209,00 | 9,053E-02 | 2,82E-15 | 1,430E-10 | 2,56E-21 | 6,46E-22 | 1,3714E-11 | 297% | 6,43E-16 | 1,18E-15 | 8,72E-16 | 8,28E-14 | 8,36E-14 | 6,04E-08 | 2,48E-09 | 3,8889E-10 | 1,0000E+00 | 0,86 |
| Po | 84 | 125 | 210,00 | 9,310E-02 | 2,82E-15 | 1,350E-10 | 2,49E-21 | 6,35E-22 | 1,3335E-11 | 292% | 6,40E-16 | 1,19E-15 | 8,72E-16 | 8,32E-14 | 8,40E-14 | 6,76E-08 | 2,79E-09 | 4,3635E-10 | 1,0000E+00 | 0,86 |
| At | 85 | 125 | 210,00 | 9,570E-02 | 2,82E-15 | 1,270E-10 | 2,41E-21 | 6,21E-22 | 1,2973E-11 | 288% | 6,33E-16 | 1,20E-15 | 8,72E-16 | 8,32E-14 | 8,40E-14 | 7,60E-08 | 3,15E-09 | 4,9305E-10 | 1,0000E+00 | 0,86 |
| Rn | 86 | 136 | 222,00 | 9,840E-02 | 2,82E-15 | 1,200E-10 | 2,35E-21 | 6,08E-22 | 1,2617E-11 | 286% | 6,31E-16 | 1,21E-15 | 8,75E-16 | 8,83E-14 | 8,88E-14 | 8,72E-08 | 3,67E-09 | 5,5225E-10 | 1,0000E+00 | 0,87 |
| Fr | 87 | 136 | 223,00 | 1,010E-01 | 2,82E-15 | 2,700E-10 | 2,26E-21 | 5,96E-22 | 1,2292E-11 | 278% | 6,18E-16 | 1,24E-15 | 8,76E-16 | 8,88E-14 | 8,92E-14 | 1,86E-08 | 7,26E-10 | 1,0909E-10 | 1,0000E+00 | 0,87 |
| Ra | 88 | 138 | 226,00 | 1,040E-01 | 2,82E-15 | 2,150E-10 | 2,19E-21 | 5,85E-22 | 1,1938E-11 | 274% | 6,14E-16 | 1,25E-15 | 8,77E-16 | 9,00E-14 | 9,04E-14 | 2,90E-08 | 1,16E-09 | 1,7204E-10 | 1,0000E+00 | 0,87 |
| Ac | 89 | 138 | 227,00 | 1,068E-01 | 2,82E-15 | 1,950E-10 | 2,12E-21 | 5,77E-22 | 1,1629E-11 | 267% | 6,06E-16 | 1,27E-15 | 8,77E-16 | 9,05E-14 | 9,08E-14 | 3,50E-08 | 1,41E-09 | 2,0914E-10 | 1,0000E+00 | 0,87 |
| Th | 90 | 142 | 232,04 | 1,096E-01 | 2,82E-15 | 1,790E-10 | 2,06E-21 | 5,66E-22 | 1,1328E-11 | 263% | 6,00E-16 | 1,28E-15 | 8,78E-16 | 9,26E-14 | 9,28E-14 | 4,17E-08 | 1,70E-09 | 2,4819E-10 | 1,0000E+00 | 0,87 |
| Pa | 91 | 140 | 231,04 | 1,126E-01 | 2,82E-15 | 1,630E-10 | 1,99E-21 | 5,56E-22 | 1,1026E-11 | 259% | 5,95E-16 | 1,29E-15 | 8,78E-16 | 9,22E-14 | 9,24E-14 | 4,97E-08 | 2,04E-09 | 2,9931E-10 | 1,0000E+00 | 0,87 |
| U | 92 | 146 | 238,03 | 1,160E-01 | 2,82E-15 | 1,380E-10 | 1,93E-21 | 5,45E-22 | 1,0703E-11 | 254% | 5,90E-16 | 1,31E-15 | 8,80E-16 | 9,52E-14 | 9,52E-14 | 6,94E-08 | 2,90E-09 | 4,1758E-10 | 1,0000E+00 | 0,87 |

Table 1. The table shows all calculations using NIST data values.

| Element | A | Entropy | | | | | |
|---|---|---|---|---|---|---|---|
| | A | S(0,83) | S(0,84) | S(0,85) | S(0,86) | S(0,87) | S(0,88) |
| Na | 22,99 | 8,76E-09 | 8,84E-09 | 8,93E-09 | 9,00E-09 | 9,10E-09 | 9,19E-09 |
| Mg | 24,31 | 1,49E-08 | 1,51E-08 | 1,52E-08 | 1,53E-08 | 1,55E-08 | 1,57E-08 |
| Al | 26,98 | 1,57E-08 | 1,59E-08 | 1,61E-08 | 1,62E-08 | 1,64E-08 | 1,66E-08 |
| Si | 28,09 | 2,59E-08 | 2,61E-08 | 2,64E-08 | 2,66E-08 | 2,70E-08 | 2,72E-08 |
| P | 30,97 | 3,38E-08 | 3,41E-08 | 3,45E-08 | 3,48E-08 | 3,52E-08 | 3,56E-08 |
| S | 32,06 | 4,19E-08 | 4,23E-08 | 4,28E-08 | 4,32E-08 | 4,37E-08 | 4,42E-08 |
| Cl | 35,45 | 5,30E-08 | 5,36E-08 | 5,42E-08 | 5,48E-08 | 5,55E-08 | 5,61E-08 |
| K | 39,10 | 6,38E-09 | 6,46E-09 | 6,53E-09 | 6,60E-09 | 6,69E-09 | 6,76E-09 |
| Ca | 40,08 | 9,91E-09 | 1,00E-08 | 1,01E-08 | 1,02E-08 | 1,04E-08 | 1,05E-08 |
| Sc | 44,96 | 1,14E-08 | 1,15E-08 | 1,17E-08 | 1,18E-08 | 1,19E-08 | 1,21E-08 |
| Ti | 47,88 | 1,27E-08 | 1,28E-08 | 1,30E-08 | 1,31E-08 | 1,33E-08 | 1,35E-08 |
| V | 50,94 | 1,37E-08 | 1,38E-08 | 1,40E-08 | 1,42E-08 | 1,44E-08 | 1,45E-08 |
| Cr | 51,99 | 1,46E-08 | 1,47E-08 | 1,49E-08 | 1,51E-08 | 1,53E-08 | 1,55E-08 |
| Mn | 54,94 | 1,57E-08 | 1,59E-08 | 1,61E-08 | 1,63E-08 | 1,66E-08 | 1,68E-08 |
| Fe | 55,85 | 1,69E-08 | 1,71E-08 | 1,73E-08 | 1,75E-08 | 1,77E-08 | 1,80E-08 |
| Co | 58,93 | 1,80E-08 | 1,82E-08 | 1,85E-08 | 1,87E-08 | 1,90E-08 | 1,92E-08 |
| Ni | 58,69 | 1,87E-08 | 1,89E-08 | 1,92E-08 | 1,94E-08 | 1,97E-08 | 1,99E-08 |
| Cu | 63,55 | 2,03E-08 | 2,05E-08 | 2,08E-08 | 2,10E-08 | 2,14E-08 | 2,16E-08 |
| Zn | 65,41 | 2,13E-08 | 2,16E-08 | 2,19E-08 | 2,21E-08 | 2,25E-08 | 2,28E-08 |
| Ga | 69,72 | 2,40E-08 | 2,44E-08 | 2,47E-08 | 2,50E-08 | 2,54E-08 | 2,57E-08 |
| Ge | 72,64 | 1,95E-08 | 1,97E-08 | 2,00E-08 | 2,02E-08 | 2,06E-08 | 2,08E-08 |
| As | 74,92 | 3,41E-08 | 3,46E-08 | 3,51E-08 | 3,55E-08 | 3,60E-08 | 3,65E-08 |
| Se | 78,96 | 4,22E-08 | 4,28E-08 | 4,34E-08 | 4,40E-08 | 4,47E-08 | 4,53E-08 |
| Br | 79,90 | 2,53E-08 | 2,57E-08 | 2,60E-08 | 2,64E-08 | 2,68E-08 | 2,72E-08 |
| Kr | 83,80 | 1,34E-08 | 1,36E-08 | 1,38E-08 | 1,40E-08 | 1,42E-08 | 1,44E-08 |
| Rb | 85,47 | 7,16E-09 | 7,26E-09 | 7,36E-09 | 7,46E-09 | 7,58E-09 | 7,68E-09 |
| Sr | 87,62 | 1,04E-08 | 1,06E-08 | 1,07E-08 | 1,08E-08 | 1,10E-08 | 1,12E-08 |
| Y | 88,91 | 1,11E-08 | 1,13E-08 | 1,15E-08 | 1,16E-08 | 1,18E-08 | 1,20E-08 |
| Zr | 91,22 | 1,19E-08 | 1,21E-08 | 1,22E-08 | 1,24E-08 | 1,26E-08 | 1,28E-08 |
| Nb | 92,91 | 1,29E-08 | 1,31E-08 | 1,33E-08 | 1,35E-08 | 1,37E-08 | 1,39E-08 |
| Mo | 95,94 | 1,42E-08 | 1,44E-08 | 1,46E-08 | 1,48E-08 | 1,50E-08 | 1,52E-08 |
| Tc | 98,91 | 1,54E-08 | 1,56E-08 | 1,59E-08 | 1,61E-08 | 1,64E-08 | 1,66E-08 |
| Ru | 101,07 | 7,66E-09 | 7,77E-09 | 7,89E-09 | 7,99E-09 | 8,13E-09 | 8,25E-09 |
| Rh | 102,91 | 2,84E-08 | 2,89E-08 | 2,93E-08 | 2,97E-08 | 3,02E-08 | 3,06E-08 |
| Pd | 107,87 | 1,86E-08 | 1,89E-08 | 1,92E-08 | 1,94E-08 | 1,98E-08 | 2,01E-08 |
| Ag | 107,90 | 1,95E-08 | 1,98E-08 | 2,01E-08 | 2,03E-08 | 2,07E-08 | 2,10E-08 |
| Cd | 112,41 | 2,08E-08 | 2,11E-08 | 2,14E-08 | 2,17E-08 | 2,21E-08 | 2,24E-08 |
| In | 114,80 | 2,23E-08 | 2,26E-08 | 2,29E-08 | 2,32E-08 | 2,36E-08 | 2,40E-08 |
| Sn | 118,71 | 2,60E-08 | 2,64E-08 | 2,68E-08 | 2,71E-08 | 2,76E-08 | 2,80E-08 |
| Sb | 121,80 | 3,10E-08 | 3,14E-08 | 3,19E-08 | 3,23E-08 | 3,29E-08 | 3,34E-08 |
| Te | 127,60 | 3,67E-08 | 3,72E-08 | 3,78E-08 | 3,83E-08 | 3,90E-08 | 3,96E-08 |
| I | 126,90 | 4,16E-08 | 4,22E-08 | 4,29E-08 | 4,35E-08 | 4,42E-08 | 4,49E-08 |
| Xe | 131,30 | 4,76E-08 | 4,83E-08 | 4,91E-08 | 4,97E-08 | 5,06E-08 | 5,14E-08 |
| Cs | 132,90 | 6,87E-09 | 6,98E-09 | 7,09E-09 | 7,19E-09 | 7,32E-09 | 7,43E-09 |
| Ba | 137,30 | 9,54E-09 | 9,69E-09 | 9,85E-09 | 9,98E-09 | 1,02E-08 | 1,03E-08 |
| La | 138,90 | 1,58E-08 | 1,60E-08 | 1,63E-08 | 1,65E-08 | 1,68E-08 | 1,71E-08 |
| Ce | 140,10 | 7,03E-09 | 7,15E-09 | 7,26E-09 | 7,36E-09 | 7,49E-09 | 7,61E-09 |
| Pr | 140,90 | 1,01E-08 | 1,03E-08 | 1,04E-08 | 1,06E-08 | 1,08E-08 | 1,09E-08 |
| Nd | 144,24 | 1,44E-08 | 1,47E-08 | 1,49E-08 | 1,51E-08 | 1,54E-08 | 1,56E-08 |
| Pm | 145,00 | 1,46E-08 | 1,48E-08 | 1,51E-08 | 1,53E-08 | 1,56E-08 | 1,58E-08 |
| Sm | 150,36 | 9,50E-09 | 9,65E-09 | 9,81E-09 | 9,94E-09 | 1,01E-08 | 1,03E-08 |
| Eu | 152,00 | 1,19E-08 | 1,21E-08 | 1,23E-08 | 1,24E-08 | 1,27E-08 | 1,29E-08 |
| Gd | 157,30 | 1,19E-08 | 1,21E-08 | 1,23E-08 | 1,24E-08 | 1,27E-08 | 1,29E-08 |
| Tb | 158,90 | 2,06E-08 | 2,09E-08 | 2,13E-08 | 2,16E-08 | 2,20E-08 | 2,23E-08 |
| Dy | 162,50 | 2,04E-08 | 2,07E-08 | 2,10E-08 | 2,13E-08 | 2,17E-08 | 2,21E-08 |
| Ho | 164,93 | 1,08E-08 | 1,10E-08 | 1,12E-08 | 1,14E-08 | 1,16E-08 | 1,18E-08 |
| Er | 167,26 | 1,29E-08 | 1,31E-08 | 1,34E-08 | 1,36E-08 | 1,38E-08 | 1,40E-08 |
| Tm | 204,38 | 2,45E-08 | 2,49E-08 | 2,53E-08 | 2,57E-08 | 2,62E-08 | 2,66E-08 |
| Yb | 173,05 | 1,36E-08 | 1,38E-08 | 1,40E-08 | 1,42E-08 | 1,45E-08 | 1,47E-08 |
| Lu | 175,00 | 1,43E-08 | 1,45E-08 | 1,47E-08 | 1,50E-08 | 1,52E-08 | 1,55E-08 |
| Ta | 180,90 | 1,69E-08 | 1,72E-08 | 1,75E-08 | 1,78E-08 | 1,81E-08 | 1,84E-08 |
| W | 183,84 | 1,83E-08 | 1,86E-08 | 1,89E-08 | 1,92E-08 | 1,95E-08 | 1,98E-08 |
| Re | 186,21 | 3,54E-08 | 3,59E-08 | 3,66E-08 | 3,71E-08 | 3,78E-08 | 3,84E-08 |
| Os | 190,23 | 2,01E-08 | 2,05E-08 | 2,08E-08 | 2,11E-08 | 2,15E-08 | 2,19E-08 |
| Ir | 192,22 | 3,59E-08 | 3,65E-08 | 3,71E-08 | 3,76E-08 | 3,83E-08 | 3,90E-08 |
| Pt | 195,08 | 3,51E-08 | 3,57E-08 | 3,63E-08 | 3,69E-08 | 3,76E-08 | 3,82E-08 |
| Au | 197,00 | 2,30E-08 | 2,34E-08 | 2,38E-08 | 2,41E-08 | 2,46E-08 | 2,50E-08 |
| Hg | 200,59 | 2,40E-08 | 2,44E-08 | 2,48E-08 | 2,52E-08 | 2,56E-08 | 2,61E-08 |
| Tl | 204,40 | 2,45E-08 | 2,49E-08 | 2,53E-08 | 2,57E-08 | 2,62E-08 | 2,66E-08 |
| Pb | 207,20 | 2,97E-08 | 3,02E-08 | 3,08E-08 | 3,12E-08 | 3,18E-08 | 3,23E-08 |
| Bi | 209,00 | 3,44E-08 | 3,50E-08 | 3,56E-08 | 3,61E-08 | 3,68E-08 | 3,74E-08 |
| Po | 210,00 | 3,85E-08 | 3,91E-08 | 3,98E-08 | 4,04E-08 | 4,12E-08 | 4,18E-08 |
| At | 210,00 | 4,32E-08 | 4,40E-08 | 4,47E-08 | 4,54E-08 | 4,62E-08 | 4,70E-08 |
| Rn | 222,00 | 4,94E-08 | 5,03E-08 | 5,12E-08 | 5,19E-08 | 5,29E-08 | 5,38E-08 |
| Fr | 223,00 | 1,05E-08 | 1,07E-08 | 1,09E-08 | 1,11E-08 | 1,13E-08 | 1,15E-08 |
| Ra | 226,00 | 1,64E-08 | 1,67E-08 | 1,70E-08 | 1,72E-08 | 1,75E-08 | 1,78E-08 |
| Ac | 227,00 | 1,98E-08 | 2,01E-08 | 2,05E-08 | 2,08E-08 | 2,12E-08 | 2,15E-08 |
| Th | 232,04 | 2,35E-08 | 2,40E-08 | 2,44E-08 | 2,47E-08 | 2,52E-08 | 2,56E-08 |
| Pa | 231,04 | 2,81E-08 | 2,86E-08 | 2,91E-08 | 2,95E-08 | 3,01E-08 | 3,06E-08 |
| U | 238,03 | 3,92E-08 | 3,98E-08 | 4,05E-08 | 4,11E-08 | 4,20E-08 | 4,27E-08 |

Table 2. In this Table we have calculated the values for minimum entropy using the proton radius values (0.83 fm ≤ $r_p$ ≤ 0.88 fm). This was done for atoms with atomic number Z=11 to Z=92 and photon energies 1.0721E-03 MeV to 1.16E-01 MeV.